\documentstyle[psfig]{aa}
\begin{document}
\thesaurus {11.05.2, 11.07.1, 11.09.2, 11.16.1, 11.16.2, 11.19.2}

\title{ Spiral galaxies with large optical warps }
\author{Vladimir Reshetnikov\inst{1,2}, Fran\c coise Combes\inst{2}}    

\offprints{F.~Combes \hfill\break(e-mail: bottaro@obspm.fr)}   

\institute{Astronomical Institute of St.Petersburg State University,     
   198904 St.Petersburg, Russia     
\and
  DEMIRM, Observatoire de Paris, 61 Av. de l'Observatoire,     
 F-75014 Paris, France}  

\date{Received 1999; accepted}

\maketitle
\markboth{V. Reshetnikov \& F. Combes:
Optical warps }{}

\begin{abstract}
 As a result of our statistical study of 540 edge-on galaxies, we present here
the images and preliminary statistical analysis
of a sub-sample of  60 galaxies, that were selected to be S-type
warped spirals. Computing the average volumic density of
galaxies from available redshift surveys, a first analysis
suggests that warped galaxies are found in denser
environments.
Only the clearest and strongest warps have been
extracted here, and therefore this sample of 60 objects gather the best
candidates for future HI or optical works on galaxy warps. 


\end{abstract}

\section{Introduction}

The majority of spiral galaxies have a warped plane, as has been revealed
in the neutral gas extended component, through HI-21cm observations 
(e.g. Bosma 1981, Briggs 1990), and in a lesser extent through optical
observations (Sanchez-Saavedra et al. 1990, Reshetnikov \& Combes 1998).
 This dynamical feature raises the problem of its origin and maintenance,
 and the numerous mechanisms that have been proposed and explored
have not yet given a definitive and satisfactory answer (e.g. the review
by Binney 1992).

Differential precession should be very quick to wrap up any warp perturbation
even in the outer parts of the galaxies (Kahn \& Woltjer 1959),
unless the potential is nearly spherical
(Tubbs \& Sanders 1979). But most warps are observed while the disk is
still a significant part of the potential, which cannot therfore be spherical.
It has been shown that coherent bending modes cannot be sustained, 
since the oscillations spectrum is continuous, for realistic disks that have no sharp
edges  (Hunter \& Toomre 1969). Models then tried to consider a non-spherical
dark halo, misaligned with the inner visible disk of the galaxy (Sparke 1984,
Sparke \& Casertano 1988, Dubinski \& Kuijken 1995). However, these 
structures can only be transient, since the inner disk is bound to align with
the dark halo (New et al. 1998, Binney et al. 1998). Alternatively, the warp 
could be the consequence of continuous accretion of gas with a slewed 
angular momentum, due to cosmic infall, as
suggested by Ostriker \& Binney (1989) and Binney (1992).  
It is not excluded either that a large part of warps are
due to interactions or mergers: the prototypical warped
galaxy NGC 5907 (Sancisi 1976) that was long thought isolated,
might have experienced a minor merger recently (Lequeux et al. 1998),
and is currently interacting with two dwarf companions  (Shang et al. 1998).

To progress about the puzzle of the origin of warps, it is important to
have a sample of optically strongly warped galaxies, to perform 
new observations and statistical studies. Recently, we have presented
a survey of optical warps in a sample of 540 galaxies, about 5 times
larger than the previous samples (Reshetnikov \& Combes 1998).
The galaxies were selected from the Flat Galaxy Catalogue of
Karachentsev et al. (1993) (FGC) and we studied their optical images
extracted from the Digitized Sky Surveys\footnote{The Digitized Sky Surveys
were produced at the Space Telescope Science Institute under 
U.S. Government grant NAG W-2166}.
We identified three classes of galaxies,
those without observable warps (30\%), and those with U-shaped (37\%) 
and S-shaped (33\%) warps. We have considered the artefacts due to 
projection effects, that could be severe in nearly edge-on galaxies,
when there are spiral arms or $m=2$ perturbations. Through numerical
simulations, it was found that the U-shape are more affected by projection effects,
but that no more than 15\% of S-shape warps could be geometrical
artefacts. On the other hand, intrinsic warps could be missed through 
projection effects (but no more than 20\%).

We therefore select a sample of 60 S-shape warped galaxies,
the strongest and clearest among the 174 found.
The selection is subjective, based on isophotal maps from the DSS.
This sample should be a  suitable material for future detailed HI and optical 
works on galaxy warps. 

\section{The sample and statistics}

Table 1 presents extraction from the Southern Extention of FGC
(Karachentsev et al. 1993, FGCE) for the galaxies with large S-type warps. 
The sample is limited by coordinates 
${\rm 0.^{h}0}~\leq~\alpha({\rm 1950})~\leq~{\rm 14.^{h}0}$,
$\delta({\rm 1950})~\leq~{\rm -17.^{o}5}$.
The columns are as
follows: galaxy FGCE, PGC and ESO number; right ascension and declination
for the epoch 1950.0; $B$ magnitude (NED\footnote{NASA/IPAC Extragalactic
Database}); heliocentric
radial velocity (NED); major and minor diameters measured on blue
films (in arcmin); morphological type; warp angle $\psi$ -- angle
measured from the galaxy centre, between the plane and average line
from centre to tips of outer isophotes (see Reshetnikov \& Combes 1998);
position angle of average line passing through the tips of outer contour
(measured from N to E) -- P.A.; direction of warp: clockwise (+) or 
counter-clockwise (--).

In the appendix, we present the DSS  images of all galaxies (in the $B_J$ passband),
rotated to horizontal.

\begin{figure}
\psfig{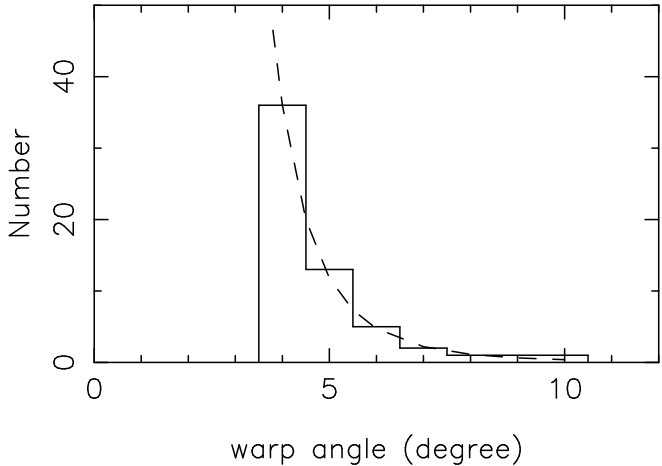}
\caption{Distribution of the observed warp angles. Dashed line
shows $\psi^{-5}$ law.}
\label{hist}
\end{figure}

Fig. \ref{hist} presents the distribution of the sample galaxies according to
warp angle $\psi$. The distribution is truncated for $\psi~\leq$~4$^{\rm o}$
since we selected only galaxies with clearest warps to avoid selection
effects. The mean value of $\psi$ is 4.$^{\rm o}$8$\pm$1.$^{\rm o}$3($\sigma$)
that is comparable with the amplitudes of optical warps found
by Sanchez-Saavedra et al. (1990), Reshetnikov (1995), and de Grijs (1997).
Dashed line in Fig.\ref{hist} shows the $\psi^{-5}$ law proposed by
Reshetnikov \& Combes (1998) to fit the observed distribution.
A naive extrapolation of this law to $\psi$=0$^{\rm o}$ suggested that
outer parts of all disk galaxies are warped with typical amplitudes
of a few degrees. 

\begin{figure}
\psfig{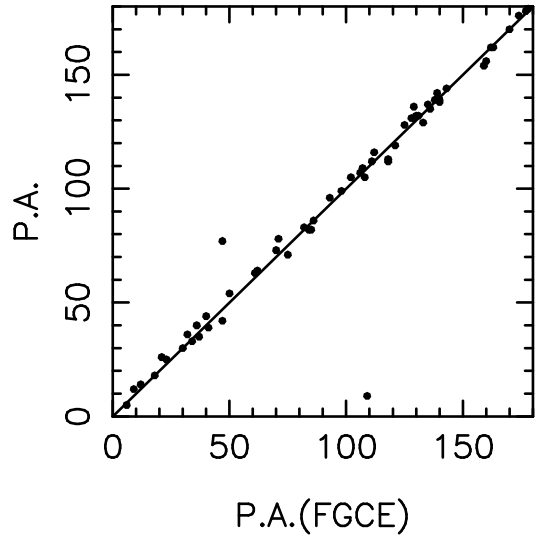}
\caption{Comparison between position angles measurements (in degrees)
in our work (P.A.) and FGCE -- P.A.(FGCE). The solid line shows equality.}
\label{pa}
\end{figure}

In Fig. \ref{pa} we compare our measurements of the position angles of the sample
galaxies with the FGCE data. The agreement is quite good. The mean difference
is $<$P.A.--P.A.(FGCE)$>$=--0.$^{\rm o}$6$\pm$1.$^{\rm o}$8(s.e.m.).
Excluding two most deviating galaxies (FGCE~333, 981) we have
$<$P.A.--P.A.(FGCE)$>$=+0.$^{\rm o}$6$\pm$0.$^{\rm o}$4(s.e.m.).

\begin{figure*}
\centerline{\psfig{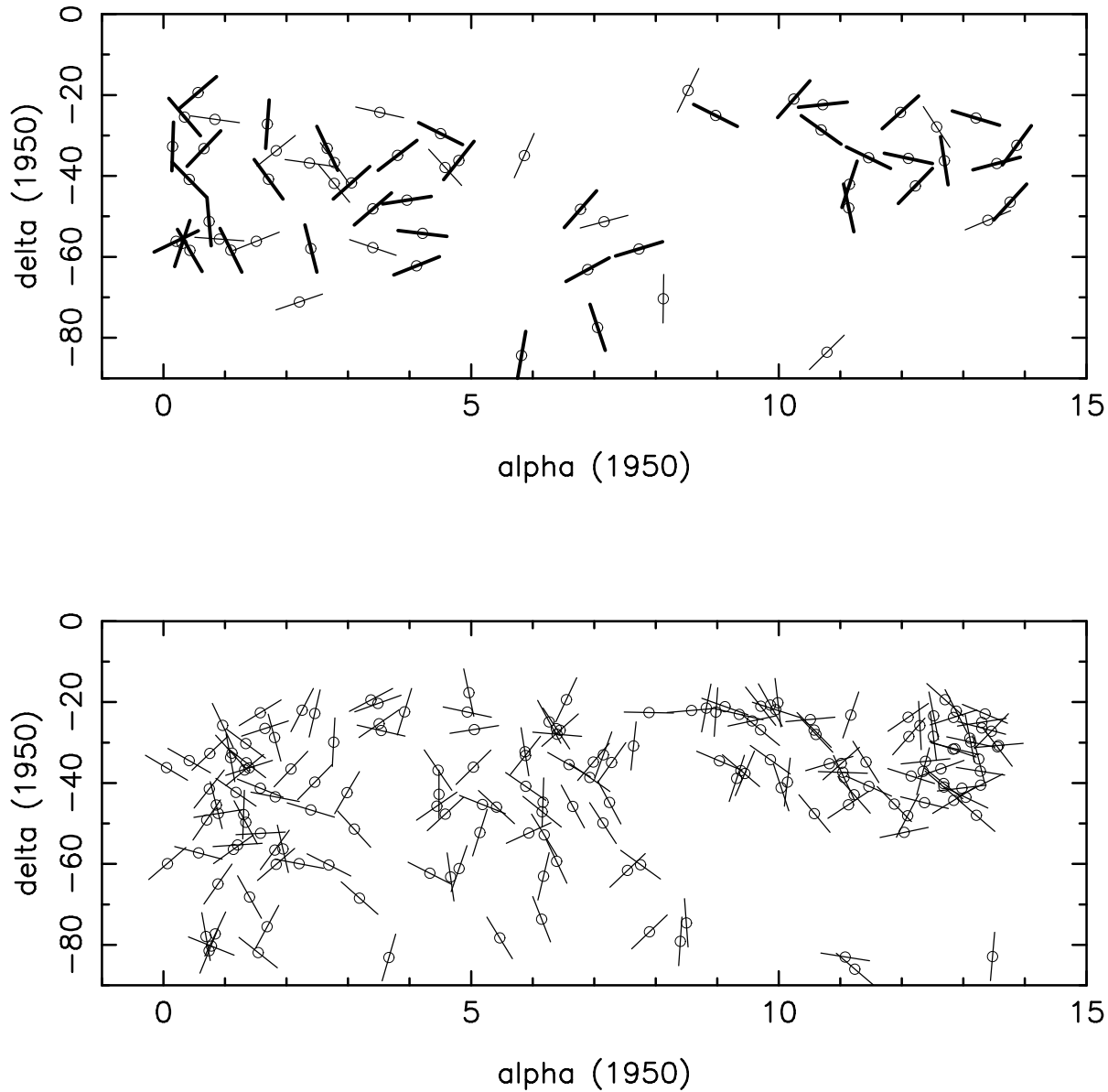}}
\caption{Distribution of the warped galaxies on the sky:
top -- S-shaped warps, bottom -- U-shaped warps.
Plots of declinations (in degrees) versus right ascension (hours).
The position angles of the galaxies (P.A.) are indicated by dashes (the 
length of each dash is 12$^{\rm o}$). Thick dashes show galaxies
with counter-clockwise warp, thin dashes -- clockwise warps
(for S-shape warped galaxies).}
\label{map}
\end{figure*}

It is evident in Figs. \ref{map} that the projected spatial distribution of
strongly warped galaxies and the distribution of their position
angles are quite homogeneous (at least in the first order approximation).
The large ``void'' in Figs.\ref{map} is due to absorption in the plane of Milky
Way. Comparison of the distributions for the galaxies with S-shaped and 
U-shaped warps shows that both distributions are statistically
undistinguishable. There is no evidence of any significant large-scale
alignment effect. 

The number of galaxies with clockwise warps (18) is smaller
than counter-clockwise galaxies (42). But, within our relatively
poor statistics, the difference is not significant (both numbers are
consistent within 3$\sigma$). 

\begin{figure}
\psfig{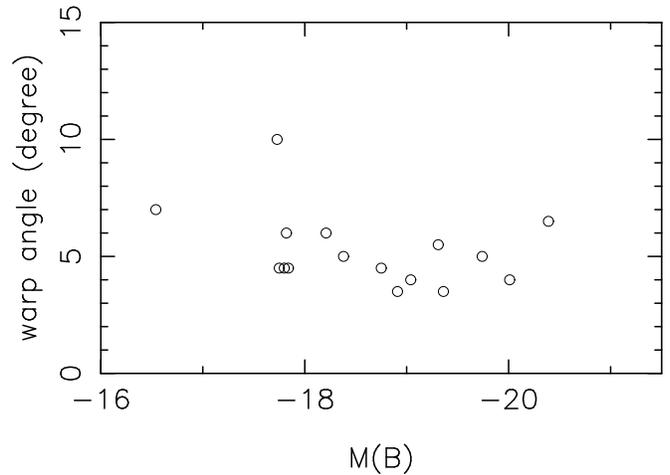}
\caption{Warp angle versus blue absolute
magnitude for the S-shape warped galaxies (H$_0$=75 km/s/Mpc).}
\label{dep}
\end{figure}

Reshetnikov (1995) found that disks of more massive and
luminous galaxies are somewhat less warped. Our present data 
do not show any significant correlation (see Fig. \ref{dep}).

\begin{table*}
\caption[1]{General characteristics of the sample galaxies}
\begin{tabular}{lllllllllllll}
\\ \\
\hline
\\
FGCE & PGC & ESO & $\alpha$(1950) & $\delta$(1950) & $B$ & V$_r$ &
$a$ & $b$ & Type & $\psi$ & P.A. & D \\
     &     &     &                &                &     & (km/s) &
(')     & (')    &     &  ($^{\rm o}$) & ($^{\rm o}$) \\ \\
\hline \\
  22 &     &  & 00 08 56.5&-32 42 50 & & & 1.18& 0.11& d&3.5&178&-- \\
  28 &  988& 149-G~024 & 00 12 21.0&-56 09 39 & 16.53 & &1.29&0.15&bc&5&116&--\\
  38 & 1324& 150-G~002 &00 18 15.0&-56 41 28 & 16.37 & &1.12 &0.15&bc&4&162&-- \\
  44 &     & & 00 20 36.8&-25 27 30 & & & 1.03 &0.13 & c&4.5&40&--  \\
  50 &     & 294-IG~011&00 24 59.7&-40 52 56 & 17.28 &&1.34&0.19&cd&4&44&-- \\
  53 & 1729& 112-G~004&00 25 43.4&-58 22 48 & 15.86& &1.49 &0.20 &c&4&30&--\\
  63 & 2167& 540-G~004&00 33 43.4&-19 24 21 & 17.08& &1.10 &0.10 &c&5&131&--  \\
  72 &     & &00 39 25.8&-33 14 36 & 15.5 & 9525 &1.05 &0.11 & bc&4&137&-- \\
  80 &     & 194-IG~037&00 44 24.4&-51 15 46 & 17.11 & &1.14 &0.11&c&8.5&5&--  \\
  99 & 3088& 474-G~035&00 50 14.9&-26 00 19 & 16.75& &1.12 &0.13&bc&4&82&+ \\
 108 & 3369& 151-G~008&00 54 27.5&-55 36 26 & 16.23& &1.23 &0.13&cd&3.5&86&+ \\
 129 & 4010& 113-G~013&01 05 45.6&-58 20 34&16.62&5914&1.25&0.15&cd&6&26&-- \\
 170 &     & &01 30 31.7&-56 07 38 & & &1.29 &0.11 & c & 4 & 112 & + \\
 187 & 6349& 477-G~001&01 41 17.0&-27 11 59 & 16.82 & &1.75 &0.11&cd&4&176&-- \\
 189 & 6394& 297-G~024&01 42 29.0&-40 49 12 & 15.91& 10171&1.26 &0.18 &bc&5&36&-- \\
 202 & 6917& 354-G~005&01 49 53.7&-33 46 34 & 15.95& 8703&1.70&0.22&bc&3.5&129&+ \\
 226 & 8499& 053-G~002&02 12 25.7&-71 08 46 & 16.21& 8000&1.88
&0.20&cd&3.5&109&+ \\
 238 & 9125& 355-G~014& 02 22 18.7&-36 47 21 & 17.37& &1.19 &0.11&c&4&82&+  \\
 240 & 9191& 115-G~011&02 23 46.2&-57 56 47 & 16.16& &1.68 &0.20 &c&5&14&--  \\
 260 &     & &02 39 39.4&-33 10 04 & & &1.01 &0.13 & dm&4.5&25&-- \\
 267 &10605& 299-G~017&02 46 22.0&-41 51 37 & 16.74&20082 &1.85 &0.22&b&6.5&39&+ \\
 269 &10640& 356-G~012&02 46 43.0&-36 43 24 & 15.82& &1.57&0.22&bc&4&35&+ \\
 294 &     & &03 03 20.5&-41 42 29 & & &1.62 &0.11 &bc&9&132&-- \\
 319 &12791& 116-G~019&03 24 07.3&-57 40 31 & 16.1& &1.23 &0.12&c&5.5&71&+  \\
 320 &     & &03 24 09.0&-48 08 40 & & &1.12 &0.11 & c&5&131&--  \\
 333 &     & 482-G~005&03 30 51.9&-24 18 06 & 15.40& 1915&1.68 &0.21&cd&7&77&+\\
 354 &13939& 359-G~001&03 48 14.8&-34 54 13 & 16.7& &1.74 &0.20 & c&3.5&128&--  \\
 363 &14212& 249-G~035&03 57 21.8&-45 59 56 & 16.24& 1031&1.90
&0.20&c&4&99&--  \\
 377 &     & &04 06 50.5&-62 10 50 & 16 & &1.08 &0.13 & cd&4&112&-- \\
 382 &14701& 157-G~010&04 12 37.5&-54 11 46 & 16.21& &1.21 &0.13 &c&4&83&--  \\
 412 &     & &04 30 10.7&-29 33 28 & & &1.14 &0.11 & c & 4 & 63 & -- \\
 416 &15621& 304-G~003&04 34 35.3&-37 53 20 & 17.43& &1.18 &0.12 &cd&3&42&+\\
 441 &16116& 361-G~012&04 48 03.9&-36 11 18 & 15.14& 5303&1.90&0.24&c&4&142&--  \\
 539 &17581& 004-G~021&05 49 10.1&-84 21 07 & 17.29& &1.12 &0.12 &c&3.5&170&-- \\
 541 &18052& 364-G~010&05 51 46.0&-34 56 36 & 16.39 & &1.19 &0.17&c&5&156&+ \\
 623 &19629& 207-G~001&06 46 41.1&-48 14 15 & 16.85& &1.01 &0.12&cd&3.5&139&-- \\
 630 &19816& 087-G~050&06 53 33.9&-63 09 19 & 15.77& 3538&1.79
&0.22&c&4.5&119&--  \\
 638 &20010& 034-G~015&07 03 21.8&-77 24 42 & 16.86& &1.34&0.12&c&5.5&18&--  \\
 642 &     &  &07 09 37.3&-51 17 38 & & 19556&1.23 &0.11 &c&5&105&+  \\
 674 &21690& 123-G~023&07 43 41.6&-58 01 52 & 14.92& 2920&2.91
&0.34&cd&4.5&107&-- \\
 690 &22797& 059-G~026&08 07 25.6&-70 20 47 & 16.81& &1.29 &0.17&b&4&179&+ \\
 706 &24027& 562-G~017&08 31 34.0&-18 52 00 & & &1.01 &0.10 &c&6&154&+  \\
 725 &25300& 496-G~025&08 58 20.3&-25 02 15 & 16.97&4677 &1.85 &0.20 &c&10&63&--  \\
 806 &30030& 567-G~038&10 14 35.3&-21 02 02 & 17.09& &1.12 &0.13 &bc&4&139&-- \\
 834 &31981& 437-G~054&10 41 17.5&-28 36 09 & 14.95& 3461&1.90 &0.27&b&5&54&--\\
 835 &32100& 569-G~003&10 42 56.4&-22 23 52 & 15.81&3731 &1.59 &0.22 &dm&4.5&96&--\\
 840 &32162& 006-G~008&10 46 54.8&-83 34 52 & 17.28& &1.57 &0.15 & c&3.5&135&+  \\
 871 &33906& 215-G~029&11 08 07.1&-47 53 40 & 16.35& &1.12 &0.16 &cd&4.5&12&-- \\
 872 &     & &11 08 57.1&-42 03 16 & & &1.01 &0.10 & c &5 &162& -- \\
 891 &     & &11 27 31.4&-35 29 25 & & &1.01 &0.09 & c &4 &64 & -- \\ \\
\hline
\end{tabular}
\label{tab1}
\end{table*}

\begin{table*}
\begin{tabular}{lllllllllllll}
\\ \\
\hline
\\
FGCE & PGC & ESO & $\alpha$(1950) & $\delta$(1950) & $B$ &V$_r$ & $a$ 
& $b$ & Type & $\psi$ & P.A. & D \\
     &     &     &                &                &     & (km/s)&
(')     & (')    &     & ($^{\rm o}$)  & ($^{\rm o}$) \\ \\
\hline \\ \\
919 &37906 & 505-G~003&11 58 32.9&-24 17 30 & 14.10 & 1808& 3.00 &0.39
&m&4.5&132&--  \\
 930 &     & &12 06 17.6&-35 39 45 & & &1.18 &0.11 & c &5&78&-- \\
 944 &39238& 321-G~017&12 13 10.2&-42 27 53 & 15.93& 6704&1.70
&0.24&b&5.5&136&--  \\
 974 &42066& 442-G~012&12 33 55.0&-27 54 00 & 16.99 & & 1.01 &0.13 & c&6&33&+\\
 981 &     & 381-G~014&12 41 26.9&-36 14 12 & 15.15& 3305&1.34 &0.18&c&6&9&--  \\
1035 &     & &13 12 15.4&-25 41 19 & & 13760&1.57 &0.11 & c &7 & 73 & -- \\
1063 &     & &13 23 55.3&-50 57 26 & & &1.01 &0.11 & c & 4.5 & 113 & + \\
1082 &     & &13 32 28.4&-36 54 31 & & &1.23 &0.11 & c &4.5&105&-- \\
1102 &     & &13 45 37.9&-46 26 26 & & &1.10 &0.13 & c&4&138&--  \\
1112 &49478& 445-G~077&13 52 12.6&-32 26 47 & 17.27& &1.03 &0.13 &c&4&144&--  \\ \\
\hline
\end{tabular}
\end{table*}

\section { The environment }

Among the warped objects, 10 galaxies are members of interacting
systems. The relative fraction of interacting galaxies -- 17\% --
is higher than the analogous fraction -- 6\% -- for our complete sample
of 540 galaxies (Reshetnikov \& Combes 1998). The fraction of isolated 
galaxies (9 objects -- 15\%) is smaller in the warped sample
than in the control sample (25\%) while the relative number of 
galaxies with companions
(68\%) is the same in both samples. This supports our conclusion
that S-shaped warps are connected with galaxy environment (Reshetnikov \&
Combes 1998). But this connection is not perfectly tight since there
are warped galaxies among relatively isolated objects
(an interpretation could be in terms of recent accretion).

\begin{figure}
\psfig{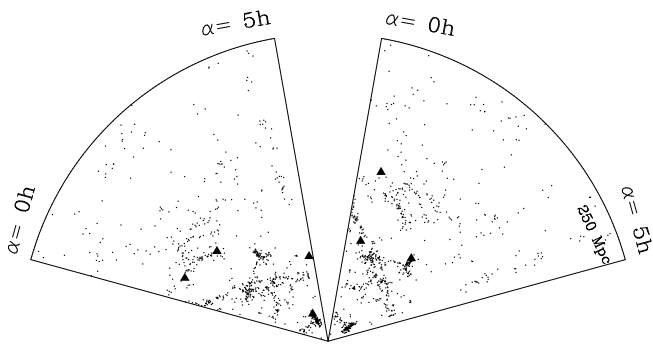}
\caption{ Location of some of the FGCE galaxies 
(filled triangles) among the SSRS2 survey objects
(dots). The left sector corresponds to warped galaxies
and right to un-warped ones, between declinations
-24 and -37$^\circ$. Right ascensions run from 
0 to 5h, and the maximum distance is 
250 Mpc (H$_0$=75 km/s/Mpc).}
\label{wed}
\end{figure}

To get more insight on the large-scale environment of warped
galaxies, we have tried to compute the average density of
galaxies around the S-shape warped population, and compare
it with a control sample. The control has been selected from
the un-warped FGCE galaxies, with the condition that the
asymmetry index along the minor axis is lower than 1.05
(cf  Reshetnikov \& Combes 1998).
We have used the Southern Sky Redshift
Survey (SSRS2, da Costa et al 1998), where redshifts and magnitudes
are reported for 5369 galaxies. Unfortunately, the redshifts
are not known for all of the FGCE catalog, and we have only 
extracted from NED 15 redshifts for the warped sample, and
17 for the control (non-warped) galaxies.
 Some of them are plotted in Fig. \ref{wed} superposed
on the SSRS2 points. 

Around each object of our sample, we compute the average 
distance of the SSRS2 galaxies, given a search radius R$_s$.
This average distance $d_m$ is computed taking the luminosity
of galaxies as weight. The mean density is then estimated as 
$$
\rho_m = { {3 N_{gal}} \over {4 \pi d_m^3} } 
$$ 
where $N_{gal}$ is the total number of objects within R$_s$.
 Taking a common weight for all galaxies only changes
 $d_m$ by 10\% at most.  The results obtained for the warped and
control samples are displayed in Table \ref{tab2}.
The average density appears 3-4 times higher for the 
warped objects. This result has to be confirmed by
more statistics, when the redshifts for the whole FGCE catalog are
known.

\begin{table}[h]
\caption[ ]{Average density around warped and non-warped galaxies}
\begin{flushleft}
\begin{tabular}{lccccc}  \hline
 R$_s$ in Mpc        & 10      &  15      &   20      &   25     &   30     \\
\hline
Warped galaxies    & 7.4E-2 & 4.2E-2 & 3.0E-2 & 1.6E-2 & 1.0E-2  \\
Un-warped sample& 1.2E-2 & 8.8E-3 & 7.3E-3 & 5.5E-3 & 4.4E-3  \\
\hline 
\end{tabular}
\end{flushleft}
R$_s$ is the search radius, and the average densities are in gal Mpc$^{-3}$ \\
\label{tab2}
\end{table}

\section{Conclusion}

In the present note we describe a new sample of southern spiral galaxies
demonstrating strong S-shape optical warps. The galaxies were selected
on the basis of their optical images from the DSS. 
First statistics indicate that warped morphologies are found
preferentially in rich environment, although this
result must be confirmed from larger redshift surveys.
The sample gives the largest available material for future 
works (optical and HI) on galaxy warps. 

\acknowledgements{
VR acknowledges support from Russian Foundation for Basic
Research (98-02-18178), ``Integration'' programme ($N$~578) and from 
French Minist\`ere de la Recherche et de la Technologie}

\section {Appendix}
 We present here a condensed summary of the 60 warped galaxies
photographs; each galaxy has been rotated by the position angle
given in Table \ref{tab1} and can be retrieved by its FGCE number.

\begin{figure*}
\psfig{figure=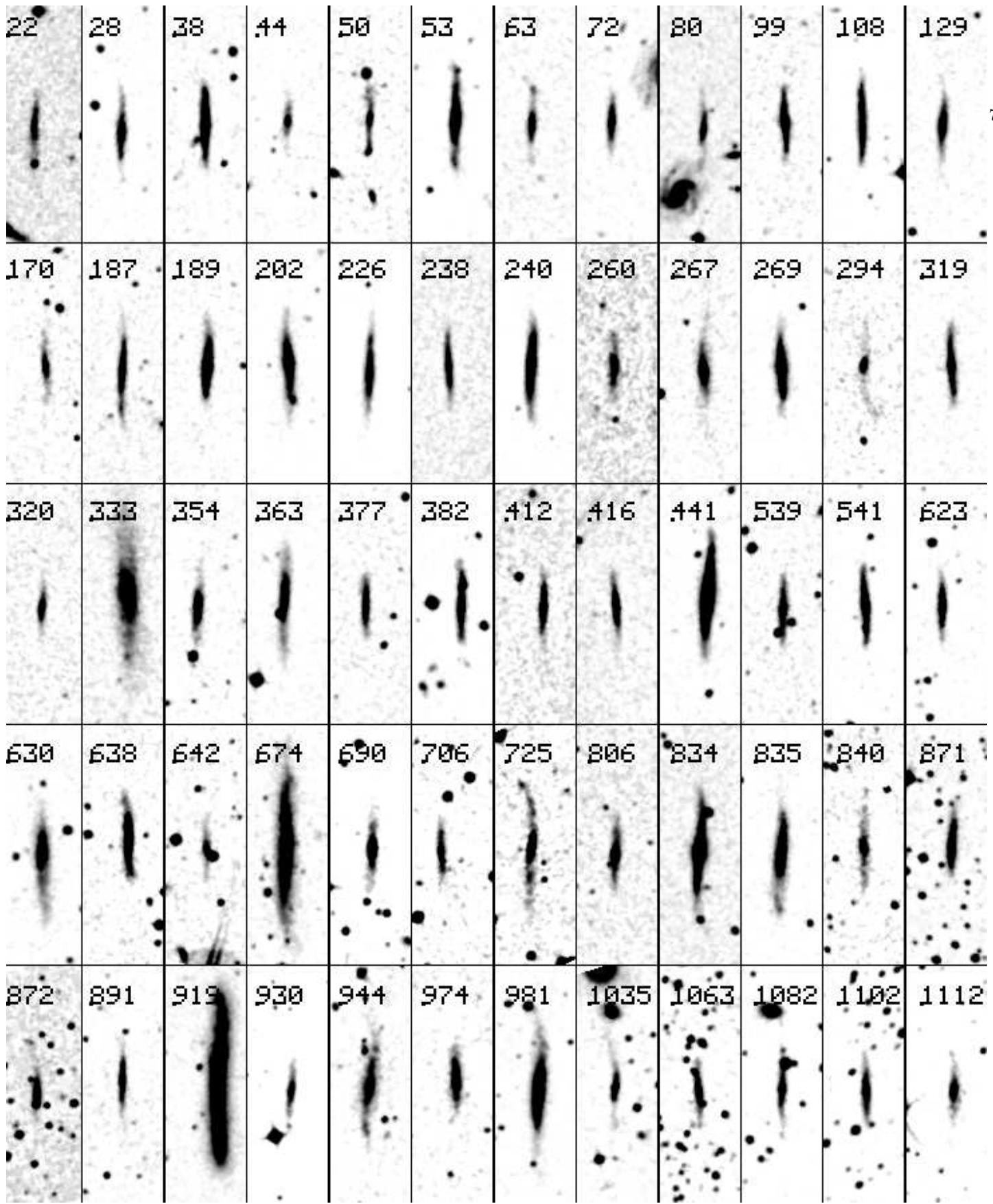,width=17cm,bbllx=17mm,bblly=25mm,bburx=197mm,bbury=248mm,angle=0}
\caption{ Digital Sky Survey images of the S-shape warped  galaxies.
The size of each image is 1'$\times$3'. The images have been rotated,
by -PA given in Table 1. }
\label{p1}
\end{figure*}

\end{document}